\documentclass[english,twocolumn,showpacs,preprintnumbers,amsmath,amssymb]{revtex4}
\usepackage{graphics}
\usepackage{amsmath}
\usepackage{subfig}
\usepackage{color}
\usepackage{graphicx}
\begin{document}
\title{Effective Mean Field Approach to Kinetic Monte Carlo Simulations in Limit
Cycle Dynamics with Reactive and Diffusive Rewiring}
\author{E. Panagakou,$^{1,2}$ G. C. Boulougouris$^{1,3}$ and A. Provata$^1$
\thanks{\emph{Corresponding author; E-mail: aprovata@chem.demokritos.gr }}
}                     

\address{$^1$ Department of Physical Chemistry - IAMPPNM, National Center for Scientific  Research
``Demokritos'', 
  Athens, GR-15310, Greece \\
$^2$ Department of Physics, University of Athens,
 Athens, GR-15771, Greece \\
$^3$ Department of Molecular Biology and Genetics,
Democritius University,  Alexandroupolis, GR-68100, Greece}
\date{Received: date / Revised version: date}
%
\begin{abstract}
 The dynamics of complex reactive schemes is known to deviate 
from the Mean Field (MF)
theory when restricted on low dimensional spatial supports. This failure has been
attributed to the limited number of species-neighbours which are available for 
interactions. In the current study, we introduce effective reactive parameters,
which depend on the type of the spatial support and which allow for an effective
MF description. As working example the Lattice Limit Cycle dynamics is used,
restricted on a 2D square lattice with nearest neighbour interactions. We show 
that the MF steady state results are recovered when the kinetic rates are replaced
with their effective values. The same conclusion holds when reactive stochastic rewiring 
is introduced in the system via long distance reactive coupling. Instead, 
when the stochastic coupling becomes diffusive the effective parameters no longer
predict the steady state. This is attributed to the diffusion process which is an
additional factor introduced into the dynamics and is not accounted for, in the 
kinetic MF scheme.
\end{abstract}
%
\maketitle

\section{Introduction}
\label{sec:intro}

The theory of reactive dynamics has for a long time mainly been restricted to studies of the macroscopic,
phenomenological, Mean Field (MF) equations.
As a result, effects such as local interactions, spatial restrictions, defects, local stochastic effects,
etc  were often ignored or added {\it ad hoc}.
In the recent years, with the development of the Kinetic
(or Dynamic)  Monte Carlo Methods, it is possible to include in detail
these factors and to follow the system as it evolves dynamically from one state to another 
\cite{jansen:1999,nagasaka:2007,petrova:2007,dedecker:2010,alvarez:2012,farkas:2012,alas:2010}.
By generating a dynamical state-to-state trajectory it is possible to explore the entire 
state space as the system
is directed towards the steady state, while the dynamics and the steady state crucially depend 
on the dynamical history.

\par In models of chemical catalytic dynamics 
\cite{nagasaka:2007,dedecker:2010,farkas:2012,imbihl:1995,imbihl:1992,imbihl:2009}, ecological models 
\cite{blasius:1999,deneubourg:2002,kouvaris:2010},
 epidemiology \cite{kuperman:2001,zanette:2002,ferreira:2001}, 
the existence of a spatial support is crucial and may modify considerably the MF
 approximation. In addition, diffusion of species may often modify the processes 
\cite{blasius:1999,kouvaris:2010}. In all these systems
the supports present certain degrees of complexity. For example, 
 in ecological systems the different species live and interact in natural environments 
 that present differences in their structure from a sub-area to another \cite{blasius:1999,baghel:2012}. 
As it is noted in \cite{baghel:2012}, in the chemical, biological and ecological world, pattern formation is a very 
common phenomenon, that must be always taken into account in respective models, since they affect the dynamics between the species.
        \par In a classic work \cite{imbihl:1992} the $NO+CO$ reaction on $Pt$ surfaces was studied, without considering the corresponding spatial effects. The stability of the system was investigated and 
local kinetic oscillations were considered.
 In a recent investigation of the same system \cite{imbihl:2009}, it is noted that the dynamics leading
 to self-organization affects the ''spatio-temporal organization" of the chemical catalytic reactions.
	\par In \cite{blasius:1999}, synchronization among populations
 of different species which are related to each other as predators-consumers-vegetation, was studied.
 The dynamics of such systems 
 is of the limit cycle type. It was found that there is a strong relation between the populations 
synchronization and the spatio-temporal characteristics of the system. This phenomenon takes also into account the possible ``diffusive migration" of some species to a neighbouring or distant areas.
	\par In epidemiological models, the spatial characteristics of the systems are very important. 
In many studies, see e.g. \cite{kuperman:2001,zanette:2002},  it is shown
 that the structure of the social network affects the dynamics of a spreading disease \cite{kuperman:2001},
 or that targeted immunization to the individuals-sites of the network that are best connected causes the 
localization of the disease \cite{zanette:2002}.
	\par In order to study these spatiotemporal phenomena,  a ``center" dynamical system, 
the Lattice Lotka Voltera (LLV) system,  was introduced on a 2-D square lattice \cite{provata:1999}. 
It was found that the conservative ``center" dynamics reduces to local oscillations when the system is 
restricted on low dimensional supports. The attributes of the oscillations depend on the lattice size, 
the number of interacting neighbors and the general spatial restrictions. These oscillations are not
 of the limit cycle type, due to the nature of the interactions.
Later on, long-distance diffusion \cite{efimov:2008} was added in the LLV system and it was shown
 that in this case the out-of-phase local oscillations get phase synchronized and give global oscillations
 which are stable. After a critical point, a Hopf-like bifurcation happens and the system enters into 
limit cycle-like dynamics.
        \par A different model, involving a limit cycle MF dynamics on lattice was introduced in 2002,
the Lattice Limit Cycle (LLC) model \cite{llc:2002,provata:2003}. For this model it was also shown that the 
behavior in the MF level and in the KMC simulations level are different. In the MF level the 
concentrations oscillate with constant amplitude, independent of the initial conditions 
(limit cycle dynamics). When the simulations are performed on a 2-D square lattice, local, 
out of phase oscillations, are observed between different-distant subregions of the lattice. 
Again, the spatial restrictions of the system, the lattice size, etc, are strong parameters
 that affect the dynamical behavior of the concentrations.
        \par In the current study, the addition of Long-Distance Reaction (LDR)
and Long-Distance Diffusion (LDD) processes is attempted on the LLC model. The aim of the study 
is to extend the previous works, by taking into account the spatial and stochastic coupling 
among the species involved in the reaction scheme and  examine how the diffusion would alter 
the behaviour of these species and the dynamics of their concentrations. In addition, 
a posteriori effective parameters are introduced in the simulations which 
are shown to restore the MF regime only in the case of LDR. In the case of LDD the MF 
results different from 
KMC results even when the effective parameters are taken into account.
	\par In the next section the LLC model is presented where three different species 
 interact in 
a predator-prey chain, and the corresponding MF equations are recapitulated. 
To implement the model, Kinetic Monte Carlo (KMC) simulations
 are performed on a square lattice and the results are shown in Sec. \ref{sec:llc}. 
Differences between the MF and the KMC approaches are 
pointed out. In Sec. \ref{sec:effect} effective reactive parameters $p'_1$, $p'_2$, $p'_3$ are calculated 
{\it a posteriori} and the MF dynamics with effective parameters are compared successfully with the KMC results.
 In the next two sections two different mechanisms and their consequences are studied.  In Sec. \ref{sec:ldr}
 long distance reactions are added to the system, while in Sec. \ref{sec:ldd} long distance diffusion is implemented. 
In both cases the species concentration is studied using the {\it a posteriori} effective values in the MF equations. 
The control parameters in these steps are the probability of long distance reaction and the probability of long distance diffusion. The concluding results as well as suggestions for future studies are presented in the Concluding section.

\section{The Lattice Limit Cycle Model: Mean Field and Kinetic Monte Carlo Approaches}
\label{sec:llc}

The LLC model is a model which describes the interactions among three completing species, 
in a chain of predator-prey interactions \cite{llc:2002}.
Each species acts as predator for one of the other two species. 
With $X_1$, $X_2$ the two interacting species (or particles) are denoted, whereas $S$ is considered to be 
a fictitious -virtual species, representing the empty sites. 
In the model, the species live on a 2-D square lattice where each site can be occupied by a
 species $X_1$, or $X_2$ or it can be empty.
Single occupancy of all sites is only allowed at all times.
 The species $S$ represents the empty sites - vacuum states of  the lattice, which "interact"
with occupied lattice sites as it will be seen in the next paragraph.
\par Initially, the species are randomly distributed onto the lattice under given
 initial conditions. Each site can interact with its closest neighbours with given rates. 
The way that the species interact with each other is described in the following scheme:

\begin{subequations}
\begin{equation}
2X_1+2X_2 \stackrel{p_1}\rightarrow  3X_2+S  
\label{eq2a} 
\end{equation}
\begin{equation}
X_1+S \stackrel{p_2}\rightarrow  2X_1 
\label{eq2b}
\end{equation}
\begin{equation}
X_2+S \stackrel{p_3} \rightarrow  2S 
\label{eq2c}
\end{equation}
\label{eq02}
\end{subequations}

The reaction step \ref{eq2a} describes the main interaction between the species $X_1$ and $X_2$.
 It is a 4-th order nonlinear reaction.
In order for this reaction to take place two species $X_1$ and two species $X_2$ are needed. 
When these particles are found in the proximity of one another (closest neighbours) then 
with rate $p_1$, one of the two $X_1$ becomes an $X_2$ while the second $X_1$  desorbs (or dies)
leaving the site empty (S).
 The reaction step \ref{eq2b} describes how the species $X_1$ is born. 
The vacuum state interacts directly with particles (states)
 $X_1$ and $X_2$. In particular, when a particle $X_1$ is found close to an empty site $S$, 
then another $X_1$ is born with reaction rate $p_2$
and gets hosted on the empty site $S$ which changes state from $S \to X_1$. 
The reaction step \ref{eq2c} describes how the species $X_2$ dies. 
When a particle $X_2$ is found close to an empty site $S$ then
with rate $p_3$, $X_2$ dies leaving its site empty, $X_2 \to S$ . 
These interactions rates, which are given by the parameters $p_1$, $p_2$, $p_3$,
will be later translated into reaction probabilities when the simulation algorithm
will be introduced.

\subsection{The Classical Mean Field Theory}
\label{sec:MF}

In this subsection we briefly recapitulate the main attributes of the classical
MF theory describing the LLC scheme \cite{llc:2002,tsekouras:2006}. The nonlinear kinetic equations for the
species concentrations have the form:
\begin{subequations}
\label{eq03}
\begin{align}
\frac{dx_1}{dt}&=-2p_1x_1^2x_2^2+p_2x_1s \label{eq3a}\\
\frac{dx_2}{dt}&=p_1x_1^2x_2^2-p_3x_2s \label{eq3b}\\
\frac{ds}{dt}&=p_1x_1^2x_2^2-p_2x_1s+p_3x_2s. \label{eq3c}
\end{align}
\end{subequations}

where the small letters $x_1$, $x_2$ and $s$ represent the
global MF particle concentrations.
The space conservation condition
\begin{eqnarray}
 x_1+x_2+s=const
\label{eq05}
\end{eqnarray}
 is automatically satisfied.
As usual, the constant is chosen equal to unity, leading to the
interpretation of $x_1$, $x_2$ and $s$ as partial concentrations of
particles $X_1$, $X_2$ and empty sites. Using the
conservation condition Eq. \ref{eq05}, it is straightforward
to reduce the system by
eliminating the $s (=1-x_1-x_2)$ variable:
\begin{eqnarray}
\frac{dx_1}{dt}&=&-2p_1x_1^2x_2^2+p_2x_1(1-x_1-x_2)
\nonumber \\
\frac{dx_2}{dt}&=&p_1x_1^2x_2^2-p_3x_2(1-x_1-x_2)
\label{eq06}
\end{eqnarray}
This reduced system admits four steady state solutions, three of
which are trivial and correspond to full occupation of the lattice
by one of the three species. The three trivial solutions can be represented 
as state vectors in the reduced ($x_1$ and $x_2$) dimensions, namely 
$Q_1=(0,0)$  (empty lattice), 
$Q_2=(0,1)$  (lattice poisoned by $X_2$) and
$Q_3=(1,0)$  (lattice poisoned by $X_1$). In addition the system \ref{eq03}
admits a  fourth nontrivial solution with coexistence of all species
in the steady state:
\begin{eqnarray}
Q_4=\left(  \sqrt[3]{\frac{p_3^2}{p_1p_2}\left[1+K\right]}
+\sqrt[3]{\frac{p_3^2}{p_1p_2}\left[1- K\right]}, \right. \nonumber \\
\left.  \sqrt[3]{\frac{p_2^2}{8p_1p_3}\left[1+ K\right]}
+\sqrt[3]{\frac{p_2^2}{8p_1p_3}\left[1- K\right]} \right) 
\label{eq07}
\end{eqnarray}
where the constant $K$ is only a function of the three reaction rates,

$K=\sqrt{1+(2p_3+p_2)^3/(27p_1p_2p_3)}$.
\par Standard linear stability analysis indicates that the first three
trivial fixed points are saddles, while the stability of $Q_4$ 
depends on the parameter values. For certain parameter values
$Q_4$ undergoes a supercritical
Hopf bifurcation, becomes an
unstable focus and in its vicinity a stable limit cycle
appears. In this regime the concentrations 
$x_1$ and $x_2$ oscillate periodically, while the amplitude
of the oscillations depends also on the system parameters.
\par Note that although the MF description is appropriate for
describing  the main long time tendencies
of the system, it is not a suitable approach when fluctuations,
spatial restrictions and stochastic effects are considered.

\subsection{The Kinetic Monte Carlo Approach}
\label{sec:KMC}

When a detailed description of the system's dynamics is needed,
where fluctuations, spatial restrictions and stochastic effects
are considered, the ultimate description is the probabilistic Master Equation
approach. It describes the temporal evolution of the system from
one state to another in detail. In particular, if the system is
found in a finite number of states $i=1,2, ... N$ then the probability
$P(j,t)$ to
find the system in state $j$, at time  $t$, is described by the 
Master Equation as:
\begin{eqnarray}
P(j,t)=\sum_{k=1}^Nw_{kj}P(k,t-1)-\sum_{l=1}^Nw_{jl}P(j,t-1)
\label{eq08}
\end{eqnarray}
where the matrix element $w_{jk}$ represents the transition probability
from state $j$ to state $k$. The matrix $W$ with elements $w_{jk}, \>\> j,k=1\cdots N$
is called the transition probability
matrix. It is straight forward to see that the solution of the Master Equation
is easy when the number of states $N$ is small, but in the case of systems
with a large number of states, as the LLC lattice model, it
becomes impossible. An alternative method where the system samples its
state space stochastically and thus for relatively large times gives a 
good approximating solution to the Master Equation is the KMC method, which
is used in the current study. 
\par The KMC method is a discrete time method starting from an initial random
configurations of $X_1$, $X_2$ and $S$ particles on lattice and the update
is random and sequential, following the LLC reactive scheme (\ref{eq02}).
As substrate a 2D square lattice of size $L \times L$
is used and each particle interacts only
locally with its nearest neighbours in the original scheme. Later, in sections
\ref{sec:ldr} and \ref{sec:ldd}, stochastic long distance couplings will be
allowed with other, distant sites. Each lattice site, whose coordinates
are denoted as $(i,j)$, contains only 
one particle ($X_1$ or $X_2$) or is empty ($S$).
Originally, the three species are randomly distributed on the lattice with
given initial concentrations. Each Elementary Time Step (ETS) of the
KMC algorithm contains four
stages:
\begin{enumerate}
\item One site, $(i,j)$, is randomly chosen through out the lattice.
\item If the site $(i,j)$ contains an $X_2$ particle and amongst the four nearest 
neighbours one particle of type $S$ is found 
then the  site $(i,j)$
changes its state from $X_2$ to $S$ with probability $p_3$. This is the
realisation of reaction (\ref{eq2c}).
\item If the site $(i,j)$ contains an $S$ particle 
and amongst the nearest neighbours a particle of type $X_1$ is found
 then the  site $(i,j)$
changes its state from $S$ to $X_1$ with probability $p_2$. This is the
realisation of reaction (\ref{eq2b}).
\item If the site $(i,j)$ contains $X_1$, and its immediate neighbourhood
 contains one $X_1$ and two $X_2$ particles, then the original
and the neighbour $X_1$ change simultaneously to $S$ and $X_2$,
with probability $p_1$. This is the lattice realisation 
of reaction (\ref{eq2a}). 
\item If none of the above three steps is realised, the lattice remains unchanged.
\item One ETS is completed. The algorithm returns to stage (1) for
a new ETS to start.
\end{enumerate}
The indicative Monte Carlo time Step (MCS) consists of a number of elementary steps equal to
the lattice size $L^2$, namely $1 \> MCS=L^2\> ETS$. 
In one MCS step each particle has reacted once, on average. The above KMC algorithm is a
slight variance of the original one proposed in ref. \cite{llc:2002}, in order to establish
consistency between the short distance interactions introduced in \cite{llc:2002}
and the long distance ones which will be introduced in the next two sections.

\par As working parameter set
the values $p_1=0.9585$, $p_2=0.016$ and  $p_3=0.026$ will be used which are located
well inside the Hopf bifurcation region and give rise to periodic oscillations with large 
amplitudes \cite{llc:2002}. In addition the working parameter values were chosen
to be $p_i \le 1$, so that they can directly be interpreted as reaction probabilities
in the KMC scheme. In cases where the reaction rates are greater than unity they can be
transformed into probabilities in two ways: a) either by dividing each one of them 
 by the sum of all $p_i$ (i.e. $p_i \to p_i/\sum {p_j} ,\>\> i,j=1,2,3$) or b)
by diving each $p_i$ by the maximum of the 
reactive rates (i.e. $p_i \to p_i/p^{max} ,\>\> i=1,2,3$). In both cases, this
rate rescaling is equivalent to a time rescaling of the system \cite{llc:2002}.
Note that the large gap in the working set value of $p_1$ in relation to $p_2$ and $p_3$ 
balances the 
highly selective, nonlinear structure of reaction step (\ref{eq2a}).
\par
\begin{figure}[h]
\centering
\includegraphics[width=0.45\textwidth,angle=0 ]{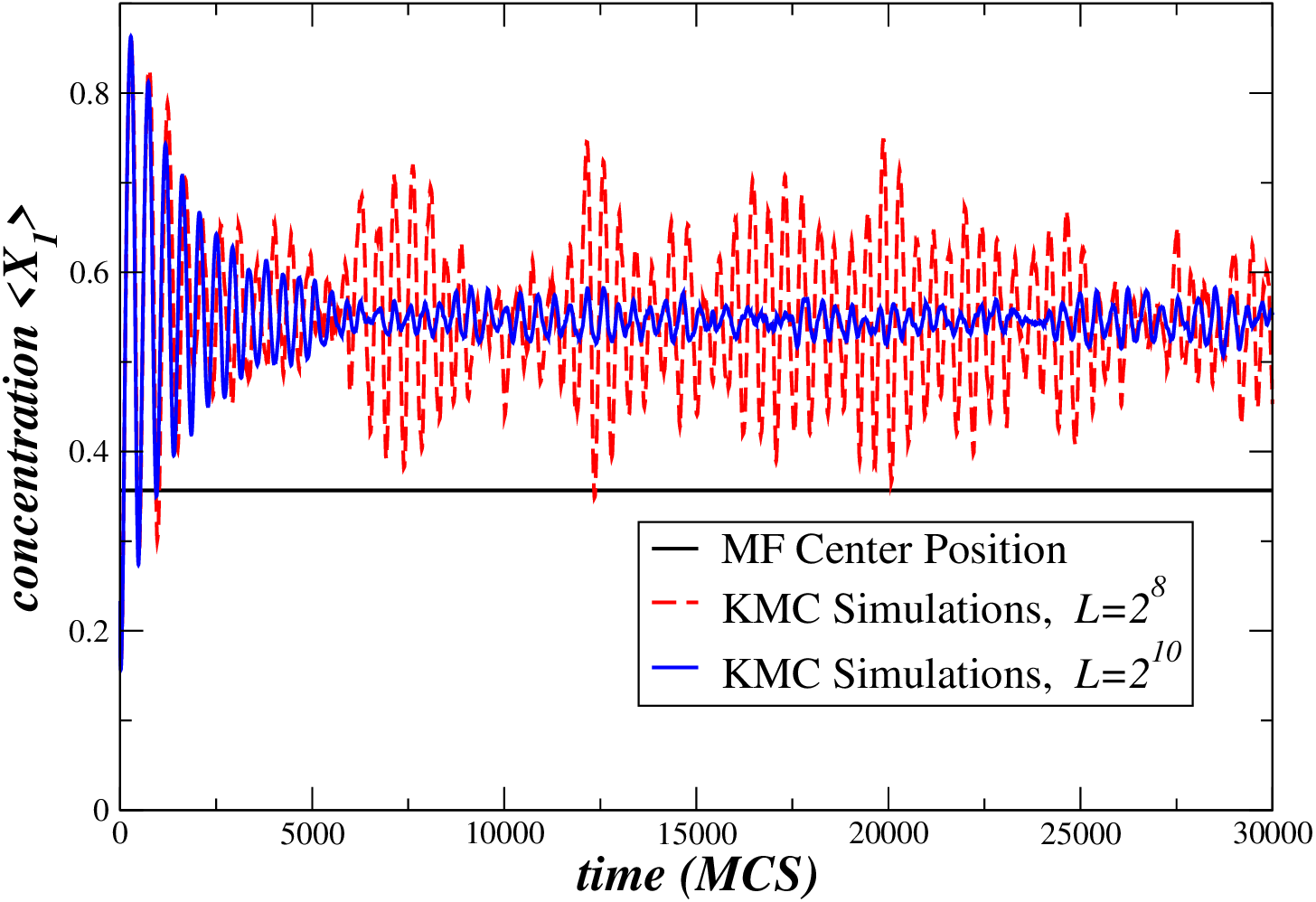}
\caption{Time series of average $X_1$ density for different lattice sizes 
shows intermittent oscillations.
The parameter values are: $p_1=0.9585$, $p_2=0.016$ and  $p_3=0.026$. The solid black line denotes the MF 
steady state solution,
Eq. \ref{eq07}.}
\label{fig01}
\end{figure}

 In Fig. \ref{fig01} the average concentration of species $X_1$ over all the lattice
is depicted as a function of time for two
different lattice sizes, $L=2^8$ (red dashed line) and $L=2^{10}$ (blue solid line).
As was also noted in \cite{llc:2002} the restriction of the reactive dynamics on a
support, the effects of spatial particle distribution and the stochastic noise induce the
following modifications to the MF behaviour: a) the oscillations loose their regularity
and become intermittent (stochastic effects), b) the amplitude of the intermittent oscillations
shrinks as the system size $L$ increases, and this is attributed to the stochastic
effects which randomise the phases of  the local limit cycle oscillators and c) a small shift is
observed in the center of the KMC cycles with respect to the MF solution. For comparison
the solid black line denotes the MF position of the center of the limit cycle. 

\section{Calculating the Effective Parameter Values}
\label{sec:effect}

\par As was shown in Sec. \ref{sec:llc}, the stochastic and spatial effects modify the reactive dynamics
with respect to the MF behaviour. Apart from the stochasticity in the amplitude of the oscillations,
the position of the center, one of the most essential characteristics of the MF, is displaced. 
Because the position
of the center is solely defined by the reactive parameters $p_1$, $p_2$ and $p_3$, it is important
to explore whether the spatial and stochastic effects influence these parameters.
\par To this purpose, we calculate, a posteriori, the ``effective'' reactive parameters, 
which realistically occur during the KMC simulations. In many cases, where a particle 
is selected for reaction with favorable rate, the event might not take place due to 
inappropriate environment (nearest neighbouring particles) or to stochastic choice.
This way the original rates are modified and new, ''effective" rates govern the dynamics
of the system. These rates are computed, a posteriori, within the algorithm provided in Sec.
\ref{sec:KMC}.
\par During the application of the algorithm we introduce three event counters $N_1$,
$N_2$ and $N_3$, which increase by one unit when the reactions \ref{eq2a}, \ref{eq2b}
and \ref{eq2c} take place, respectively. The "effective" parameters, $p'_i$, $i=1,2,3$,
are calculated as \cite{swendsen:2002}:
\begin{subequations}
\begin{align}
p'_1 & =\frac{N_1}{\left< x_1\right> ^2\left< x_2\right>^2}, \label{eq09a}\\
p'_2 & =\frac{N_2}{\left< x_1\right> \left< s\right> }, \label{eq09b}\\
p'_3 & =\frac{N_3}{\left< s\right> \left< x_2\right> } \label{eq09c}
\end{align}
\label{eq09}
\end{subequations}
\par
\begin{figure}[h]
\centering
\includegraphics[width=0.45\textwidth, angle=0]{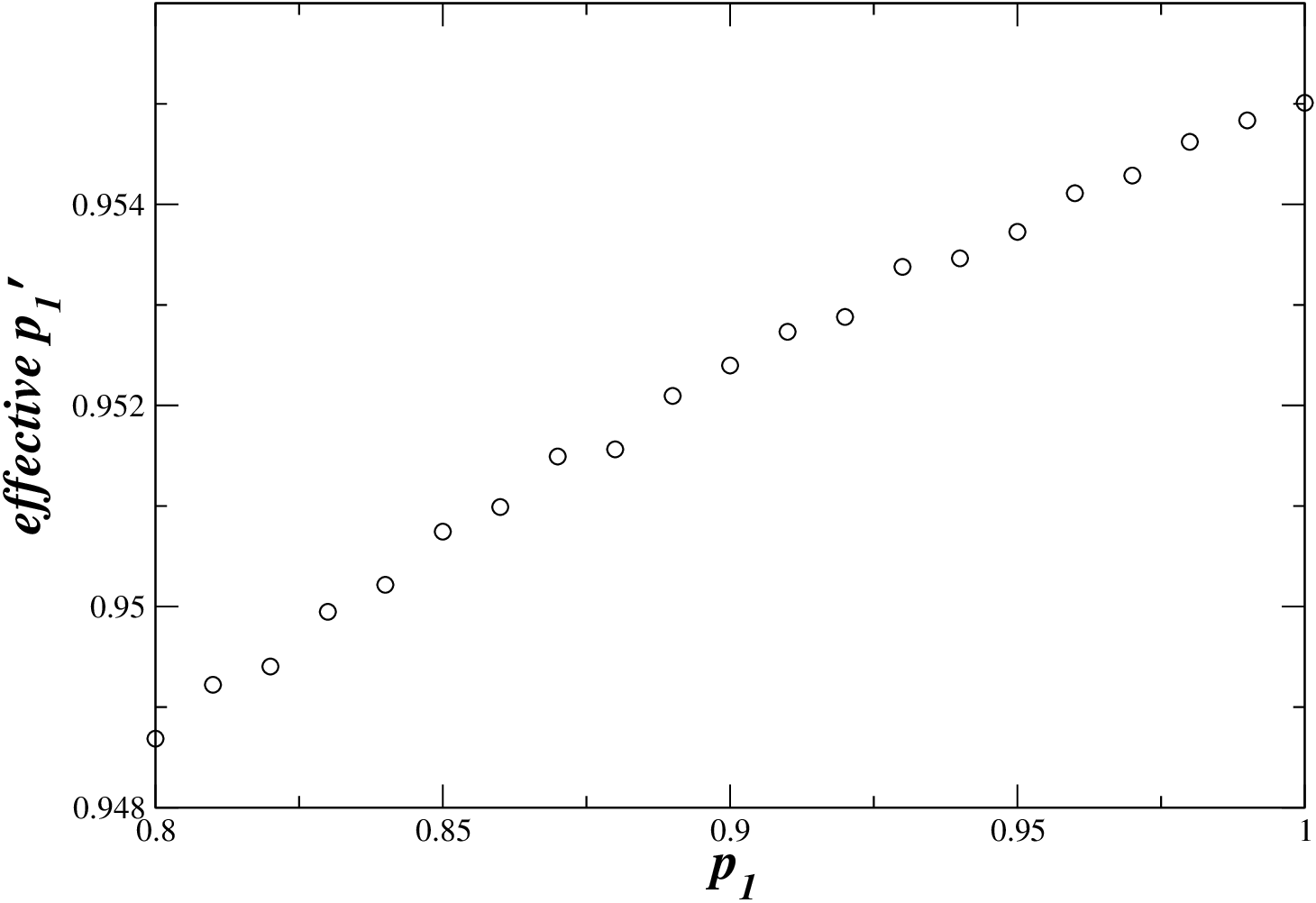}\\
\caption{Average effective reaction probabilities $p'_1$ as functions of the imposed reactive 
reactive coupling $p_1$. 
The other parameter values are: $p_2=0.016$ and  $p_3=0.026$ 
and the system size is $L=2^9$.
Averages are taken over 10 runs. The error bars on the $p'_1$ values
are of the order of $4\times 10^{-4}$.}
\label{fig02}
\end{figure}

 As an example, in Fig. \ref{fig02}, the effective values $p'_1$ is plotted for various
values of $p_1$ when the other parameters $p_2$ and $p_3$ are kept
constant. From the figure it
 is clear that as the reaction probability increases the effective reaction probability,
which averages out all the local effects,
also increases proportionally. 

\par
\begin{figure}[h]
\centering
\includegraphics[width=0.45\textwidth, angle=0]{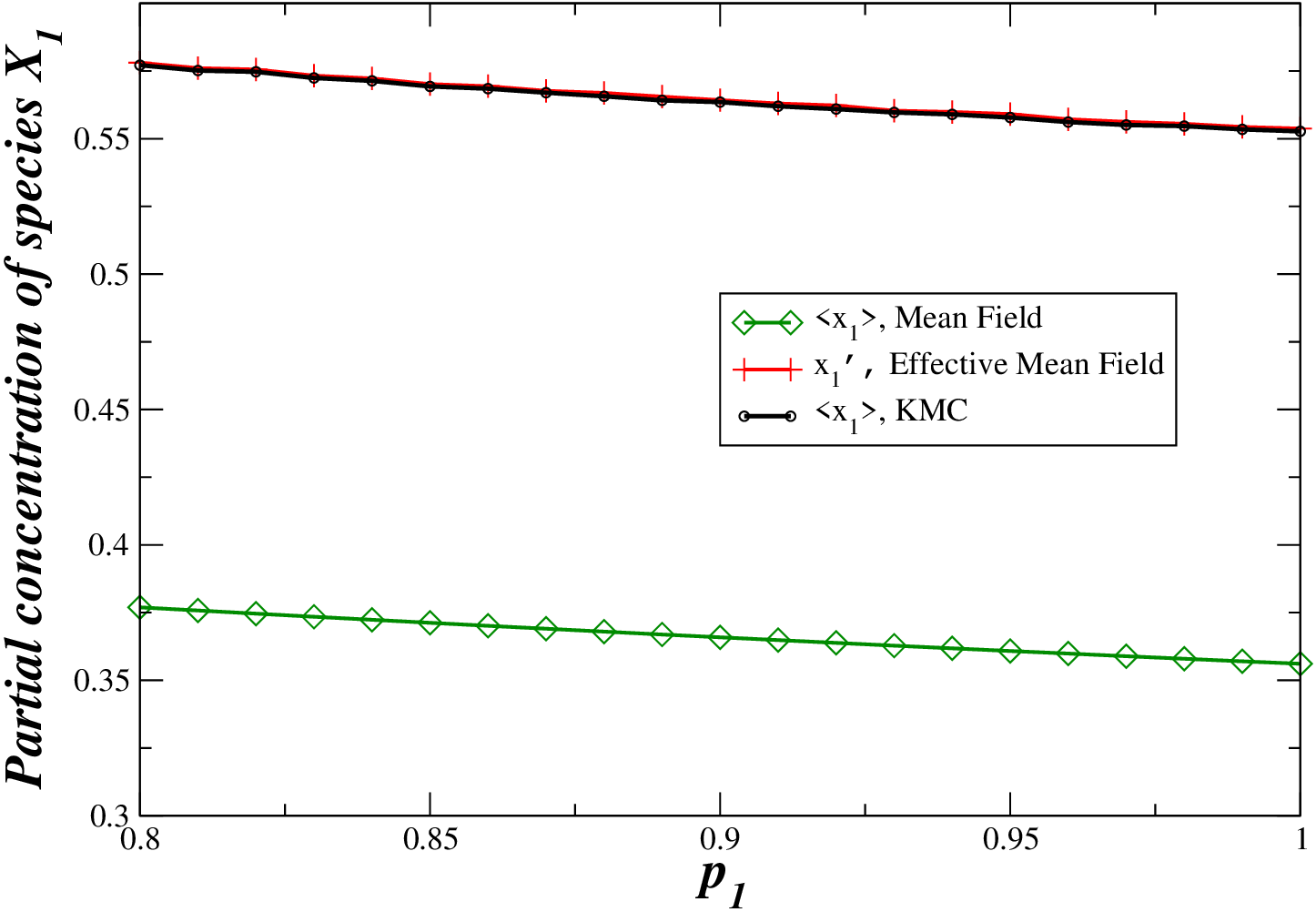}
\caption{ Effective average partial density $x'_1$ (red line with crosses) 
and KMC  average partial density $\left< x_1\right>$ (black line with circles) as a 
function of the imposed reactive 
reactive coupling $p_1$. The MF partial density $x_1$
(using the original $p_i$ values) is also plotted
for comparison.
Other parameter values are: $p_2=0.016$ and  $p_3=0.026$ 
and the system size is $L=2^9$.
KMC averages are taken over 10 runs.}
\label{fig03}
\end{figure}

 Using the effective parameters $p'_i$, the effective MF steady state values 
 $x'_1$, $x'_2$ and $s'$ are calculated using Eq. \ref{eq07}, where the kinetic
parameters are replaced with the effective ones. The results are plotted in \ref{fig03}
for various values of the original parameter $p_1$ and keeping the other two 
constant ($p_2=0.016$ and  $p_3=0.026$). In particular, in Fig. \ref{fig03} 
the three curves represent: a) the averaged values $\left< x_1\right> $
taken directly from the KMC simulations, average taken over 10 runs (black line with circles),
b)  the MF calculated effective partial density 
$x'_1$ of species $X_1$,
using the effective probabilities $p'_i$ in the MF Eq. \ref{eq07}, (red line with crosses)
and c) the original MF partial density 
$x_1$, using the original probabilities $p_i$ in the MF Eq. \ref{eq07} (green line with diamonds).
The simple MF clearly underestimates the average partial population densities, while the effective MF
gives a very close approximation to the KMC simulations. These results indicate that, as far as the
steady state properties of the LLC model are concerned, the restriction of the system on the substrate
together with the stochastic character of the reaction lead to new effective parameter values, 
while the MF steady state is achieved for these shifted parameter values. 

\section{Kinetic Lattice Monte Carlo with Stochastic Reactive Rewiring}
\label{sec:ldr}
As discussed earlier, stochastic rewiring may take place in networks where species interact with far
away "neighbours". As an example we give opinion exchange through the Internet or through the telephone
network. Opinions are discussed and exchanged without actual displacement of the individuals. In such interaction
environments the individuals are exposed both to the local environment (family, work) where the interactions
are governed by rates $p_i$, and to the long distance environment where the interactions are governed by
rates $p_{i,long}$ which, in general, are different. The long distance reactions introduce a type of 
reactive mixing in the system and drive the system towards its MF behaviour. In some cases the
distant interactions are considered using delays, but in the current study both distant and local
interactions are assumed to take place simultaneously.
\par To realise the long distance rewiring process, a rewiring rate $r$ is introduced which denotes the relative
rate for long distance versus local interactions. For the local
(short distance) reaction rates the reaction rates are denoted as $p_i$, $i=1,2,3$, as before, while for the 
long distance reaction rates the parameters are denoted as $p_{i,long}$,  $i=1,2,3$. 
\par
Having chosen the values of $r$, $p_{i}$ and $p_{i,long}$,  $i=1,2,3$ the modifications 
to the KMC algorithm are straight forward. With probability $1-r$ the classical KMC as described 
in section \ref{sec:KMC} is realised.
With probability $r$ the same algorithm holds, but now the neighbours are randomly chosen between all particles
in the system and the reactions  are realised with rates $p_{i,long}$. In one ETS either a local event or a
distant event may take place, thus the ratio of number of ETS where local interactions take place to the
number of ETS where long distance events happen is $r/(1-r)$.
\par To find out how the rewiring process modulates the local process we choose to work with the same reaction
rates, i.e. $(p_1,p_2,p_3)=(p_{1,long},p_{2,long},p_{3,long})=(0.95,0.016,0.026)$. 
In this case, while the local interactions
drive the steady state away from the MF solution, the rewiring process drives the system towards the MF because
the "nearest neighbours" are drawn with equal probability with the entire system. 
\par
\begin{figure}[h]
\centering
\includegraphics[width=0.45\textwidth, angle=0] {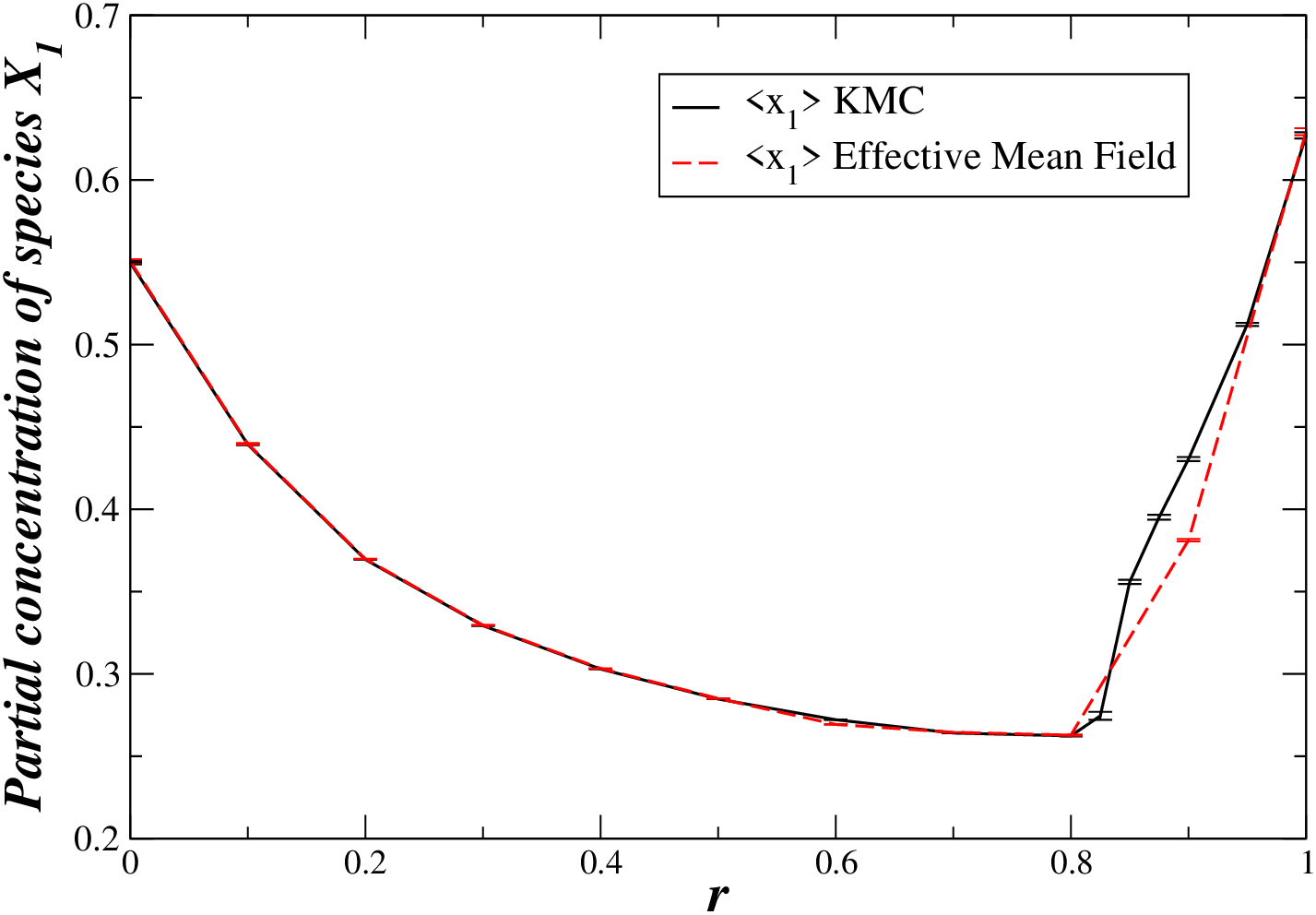}
\caption{Average densities $X_1$ as a function of the rewiring rate $r$.
 The black solid line 
denote the KMC simulations results while the red dashed line denotes the effective MF values of 
the species concentration $x_1$.
The parameter values are: $(p_1,p_2,p_3)=(0.95,0.016,0.026)$
 and the system size is $L=2^9$. Averages are taken over 10 runs and the error bars
are designated in the plot.}
\label{fig04}
\end{figure}

 In Fig. \ref{fig04} the average position of the $x_1$ variable is shown as a function of the rewiring
rate $r$. It is first noted that the introduction of a rewiring process introduces a nontrivial deviation on the
average values as a function of the rate $r$. The average partial density $x_1$ initially decreases 
as the rate of rewiring increases up to a critical rate $r_c\approx 0.8 $. After this value the
average partial density $x_1$ starts increasing abruptly. The interpretation is that for small values
of the rewiring rate $r$ the local properties dominate, while for large values of $r$ the long 
distance interactions dominate. $r_c$ stands for the specific rewiring value where the global 
properties take over. The same effects are also shown in the behaviour of the other variables
$x_2$ and $s$.

\par In the same plot the average $x_1$ densities are computed using the effective parameters $p_{i,eff}$
via Eq. \ref{eq07}. With the use of the effective parameters the average population densities are 
very well predicted by the MF equations. This comes as no surprise, after the successful results of
the previous section, where the effective MF theory predicted well the local steady state concentrations.
Here, the introduction of the long distance KMC rewiring drives the system towards MF and thus the effective
MF approximation has a higher advantage in the description of the system over the previous, purely
local KMC approach.  
\par
\begin{figure}[h]
\centering
\includegraphics[width=0.45\textwidth, angle=0] {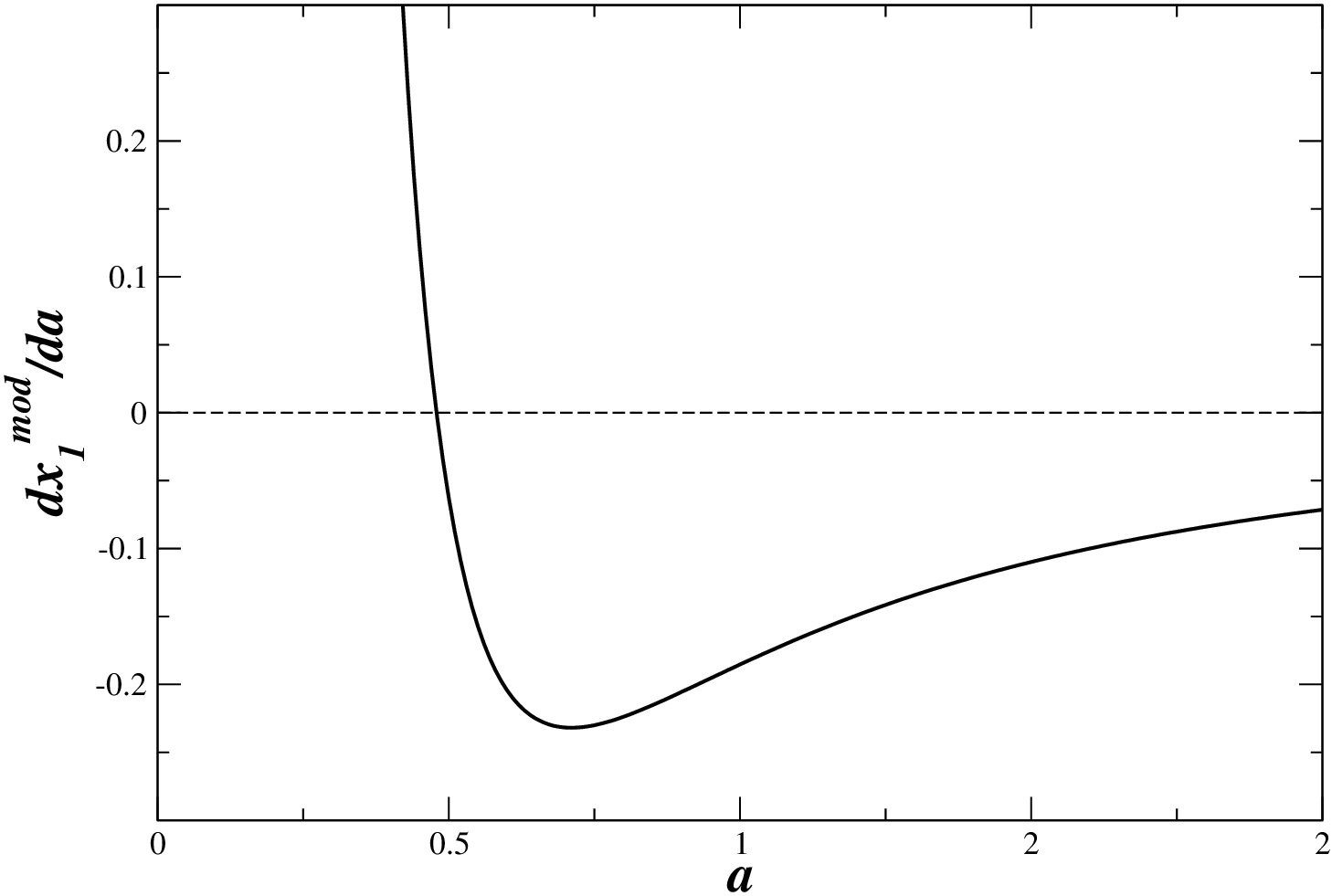}
\caption{The derivative $dx_1^{mod}/da$ as a function of $a$.}
\label{fig05}
\end{figure}

 It it possible to obtain the minimum of curve \ref{fig04} if we assume that the overall collective
 effects due to the stochastic
KMC process and to the long distance reactions is a multiplicative factor $a$, which facilitates 
the $p_1$ rate as $a^4p_1$, the $p_2$ rate as $a^2p_2$ and the $p_3$ rate as $a^2p_3$. Then the
critical value $a_c$ could be obtained as the minimum with respect to $a$
of Eq. \ref{eq07} (with the $a-$modified rates), i.e. the minimum with respect to $a$ of 

\begin{eqnarray}
Q_4^{mod}=\left( \sqrt[3]{\frac{a^4p_3^2}{a^6p_1p_2}\left[1+K^{mod}\right]}
+\sqrt[3]{\frac{a^4p_3^2}{a^6p_1p_2}\left[1- K^{mod}\right]},\right. \nonumber \\ 
\left. \sqrt[3]{\frac{a^4p_2^2}{8a^6p_1p_3}\left[1+ K^{mod}\right]}
+\sqrt[3]{\frac{a^4p_2^2}{8a^6p_1p_3}\left[1- K^{mod}\right]} \right) 
\label{eq10}
\end{eqnarray}
where $K^{mod}=\sqrt{1+(2p_3+p_2)^3/(27a^2p_1p_2p_3)}$.
For example, for the working parameter set, the minimum of the $x_1^{mod}$ curve with respect to $a$
 is attained when the derivative of the $x_1^{mod}$-component in Eq. \ref{eq10}
with respect to $a$ vanishes. In Fig. \ref{fig05} the derivative is plotted with respect to 
$a$. It crosses the $y=0$-axis at a single value $a_c\sim 0.48$, which also corresponds to the 
single minimum $r_c$ in Fig. \ref{fig03}. 

\section{Kinetic Lattice Monte Carlo with Stochastic Diffusive Rewiring}
\label{sec:ldd}
Long distance, stochastic motion (or long distance diffusion)  may take place in networks with
mobile species \cite{efimov:2008}. As an example, we give the long distance migration in ecology, especially in
bird motion.  In such systems species migrate in variable distances and they react locally
wherever they land. There are many systems in which long distance reaction and long distance
diffusion are both possible. A common one is opinion dynamics where the individuals can interact
with other distant individuals (over the phone or via the Internet) and can also relocate and
then react locally. Another such network is the sales network where potential clients can be
contacted either by Internet (long distance reaction) or by salesmen (long distance motion-diffusion).

\par In this section  we attempt to compare how the long distance reaction and long distance
diffusion mechanisms affect the output of the reactive dynamics.
Both mechanisms, long distance reaction and long distance diffusion, cause a mixing in the
system and one could assume that the two processes lead to very similar output results.
 Nonetheless, this is not the case
and, as it will be shown later in this section, the two processes combined with the local reactions
give very different overall dynamics.

\par To compare with the previous section, \ref{sec:ldr}, we only consider here local reactions together
with long distance diffusion mechanisms. The local reaction are realised according to the KMC algorithm
described in section \ref{sec:KMC}, while all the local reaction parameters are the same as in sections
\ref{sec:KMC} and \ref{sec:effect}. The local reactions are
governed by rates $p_i$, while the long distance diffusion rates are denoted by $p_{i,diff}$. 
In general the rates $p_{i,diff}$ may be different for the different species. In the current study,
in order to reduce the number of parameters we assume that all species diffuse with the same rate
$p_d$.
\par
Having chosen the values of $p_d$ and $p_{i}$,  $i=1,2,3$, the modifications to the KMC algorithm 
of section \ref{sec:KMC} are straight forward.
Once a lattice site is selected 
the hosted particle reacts according to the classical KMC as described in section \ref{sec:KMC}
with probability $1-p_d$. Otherwise,  with probability $p_d$ the hosted particle exchanges position
with another, randomly selected, particle on the lattice.
 In one ETS either a local reaction event or a long distance diffusion
event may take place, thus the ratio of number of ETS where local interactions take place to the
number of ETS where long distance events happen is $p_d/(1-p_d)$, in direct similarity with the 
long distance reaction process presented in the previous section, Sec. \ref{sec:ldr}.
In addition, for comparison with previous results we choose to work with the same local reaction
rates, i.e. $(p_1,p_2,p_3)=(0.95,0.016,0.026)$ and with variable $p_d$. The diffusion range 
$l_d$ for the
simulations presented here are set to $l_d=L/2$, but similar results are obtained for all $l_d$,
which are greater than the size of the local oscillators $R$ \cite{llc:2002} which the current 
working parameter set is $R\sim 10$.

\begin{figure}[h]
\centering
\includegraphics[width=0.45\textwidth, angle=0]{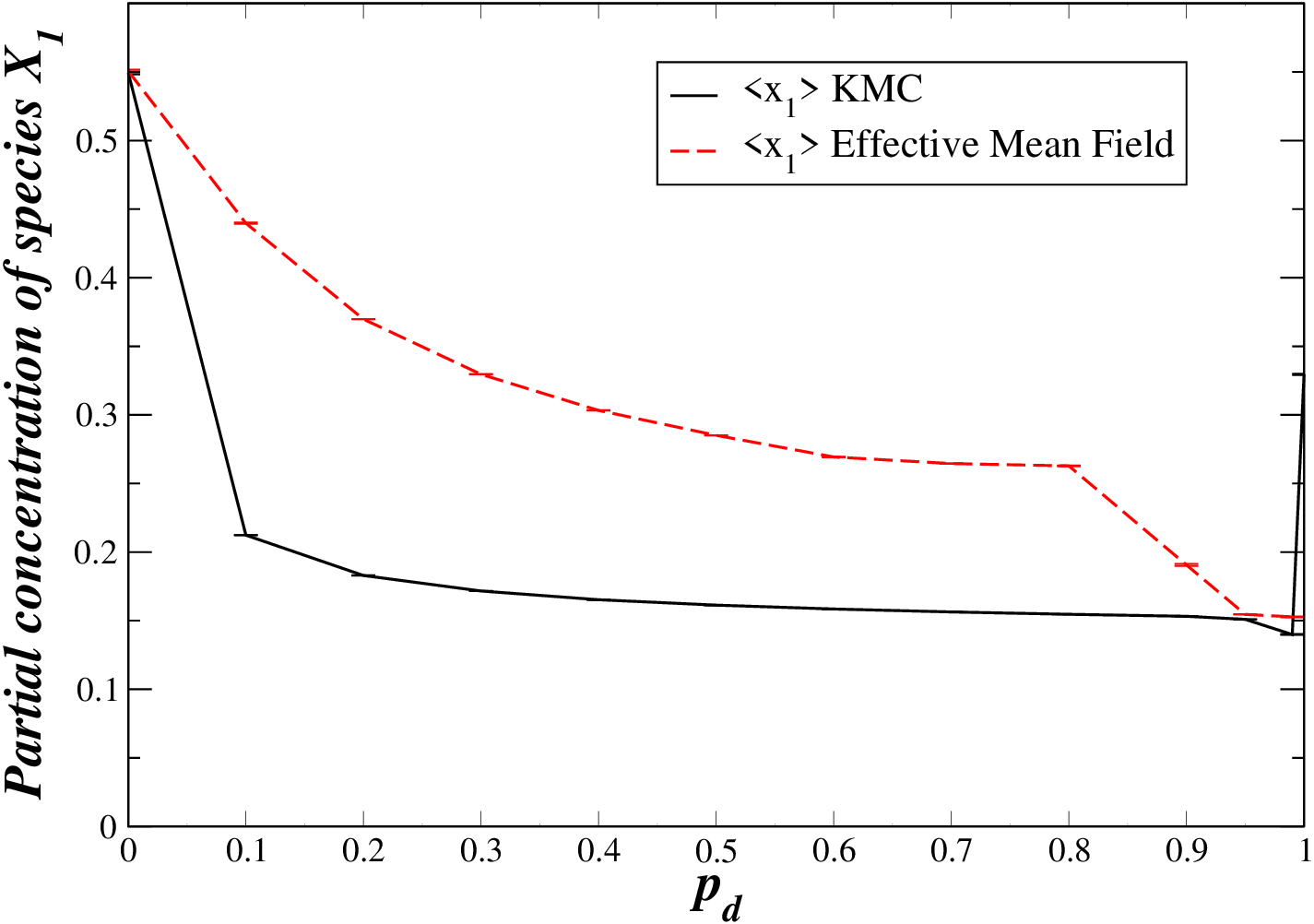}\\
\caption{Average partial densities $x_1$ as a function of the long distance reactive coupling $p_{d}$.
The black solid line denotes the KMC simulations results, and the red dashed
curve denotes the Effective MF values of $x_1$.
The parameter values are $(p_1,p_2,p_3)=(0.95,0.016,0.026)$ and the system size is $L=2^9$.
Averages are taken over 10 configurations and the error bars are shown.}
\label{fig06}
\end{figure}
\par As can be seen in Fig. \ref{fig06} there is a decrease in the rate of production
of $X_1$ particles, which takes place already for small values of diffusive mixing, while in the
case of reactive mixing the decrease is more gradual (see Fig. \ref{fig04}). When $p_d=1$
only diffusive mixing takes place, the effective reaction rates $p_{i,eff}$ are 0
 and the partial densities remain unaltered. That is, for $p_d=1$, $x_1=x_1(t=0),\>\>
x_2=x_2(t=0),\>\> s=s(t=0)$. For equiprobable initial conditions which are 
often used in the KMC process, the final states is $x_1=x_2=s=1/3$ for $p_d=1$.

\par In the same figure the red dashed line represents the predictions of the effective MF 
approach. Unlike in the case of reactive mixing the effective MF theory fails to predict
the resulting rates. This is because the introduced diffusion is a new process
which is not taken into account by the effective MF model. And although
long distance reaction and long distance diffusion both induce  mixing effects, the details
of each processes lead to different outputs in the steady state production of the three
species.

\section{Conclusions}
\label{sec6}

\par In this work we study the impact of long distance reaction and diffusion effects on the
steady state of a process governed by local Limit Cycle Dynamics. The process is realised on 
a square lattice
using Kinetic Monte Carlo simulations. 
\par For the case where only local interactions are considered, it is shown that the calculation
of {\it a posteriori} effective reaction rates, allows to get a good
approximation of the local system steady state properties, using the MF effective rates.
Further more it is shown, that when long distance reactions together with local ones
take place, the effective
MF approximation still faithfully describes the steady state properties of the system.

\par When long distance diffusion on the local reactive dynamics are added the effective MF
description is no more valid. This is because the diffusion is a new process which is not
taken into account by the original MF equations which include only reactive terms. Thus the
effective MF, based only on the reactive terms fail to describe the composite system. An extended
MF model needs to be devised which includes also diffusion effects as a better candidate on
which to base an effective MF theory to describe the dynamics of the LLC model together
with the long distant diffusion terms. 

\par In the current study, we have only discussed the properties of the steady state and not
the dynamics around it. In future studies the influence of the long distance reactive and diffusive
processes on the dynamics of the limit cycle, on synchronisation effects and the position of the
bifurcation point need to be addressed.

{\bf Acknowledgments:} The authors would like to thank Profs. A. V. Shabunin,
F. Diakonos and 
D. Frantzeskakis for helpful discussions. E.P. acknowledges
financial support from the ``Demokritos'' PhD Fellowship
Program.
This research has been co-financed by the European Union (European Social Fund – ESF) 
and Greek national funds through the Operational Program "Education and Lifelong Learning" 
of the National Strategic Reference Framework (NSRF) - Research Funding Program: THALES. 
Investing in knowledge society through the European Social Fund.


\begin{thebibliography}{99}

\bibitem{jansen:1999}    A.P.J. Jansen and J.J. Lukkien,
Catalysis Today, \textbf{ 53}, 259–271 (1999). 

\bibitem{nagasaka:2007} M. Nagasaka, H. Kondoh, I. Nakai and T. Ohta
J. Chem. Phys., \textbf{126}, 044704 (2007).

\bibitem{petrova:2007}
N. V. Petrova and I. N. Yakovkin,
European Physical Journal B \textbf{58}, 257-262 (2007). 

\bibitem{dedecker:2010}
Y. De Decker and F. Baras,
European Physical Journal B \textbf{78}, 173-186 (2010).


\bibitem{alvarez:2012} L. Alvarez-Falcon and L. Vicente,
Int. Journ. of Qunatum Chemistry \textbf{112}, 1803-1809 (2012).

\bibitem{farkas:2012}
A. Farkas, F. Hess and H. Over,
J. Phys. Chem. C \textbf{116}, 581-591 (2012). 

\bibitem{alas:2010} S.J. Alas and L Vicente,
Surface Science \textbf{604}, 957-964 (2010).  

\bibitem{imbihl:1995} R. Imbihl and G. Ertl, Chem. Rev. {\bf 95},
697 (1995).

\bibitem{imbihl:1992} R. Imbihl, T. Fink and K. Krisher, J. Chem. Phys. {\bf 96},
6236 (1992).

\bibitem{imbihl:2009} R. Imbihl, Surface Science {\bf 603} (10-12): 1671 - 1679 (2009).

\bibitem{blasius:1999} B. Blasius, A. Huppert, L. Stone. Nature {\bf 399} 354 (1999).



\bibitem{deneubourg:2002} J.L. Deneubourg, A. Lioni, C. Detrain. Biol. Bull. {\bf 202} 262 (2002).

\bibitem{kouvaris:2010} N. Kouvaris, A. Provata, D. Kugiumtzis. Phys. Lett. A {\bf 374} 507–515 (2010).

\bibitem{kuperman:2001} M. Kuperman and G. Abramson. Phys. Rev. Lett. {\bf 86}: 2909 (2001).

\bibitem{zanette:2002} D.H. Zanette and M. Kuperman. Physica A {\bf 309}: 445 (2002).

\bibitem{ferreira:2001}
C. P. Ferreira, J. F. Fontanari and R. M. Z. dos Santos,
Phys. Rev. E, \textbf{64}, 041903 (2001).

\bibitem{baghel:2012} R.S. Baghel, J. Dhar, R. Jain. E. J.,
 Diff. Equations {\bf 21} 1-12 (2012).

\bibitem{provata:1999} A. Provata, G. Nicolis and F. Baras, J. Chem.
Phys. {\bf 110}, 8361 (1999).

\bibitem{efimov:2008} A. Efimov, A. Shabunin, A. Provata,  Phys. Rev. E {\bf 78}, 056201 (2008).

\bibitem{llc:2002} 
A. V. Shabunin, F. Baras, A. Provata,
PHYSICAL REVIEW E {\textbf 66}, 036219 (2002).

\bibitem{provata:2003} A. Provata, G. A. Tsekouras, F. Diakonos, et al.,
Fluctuation and Noise Letters {\textbf 3}, L241-L250 (2003).

\bibitem{tsekouras:2006} G. A. Tsekouras, A. Provata,
European Physical Journal B {\textbf 52}, 107-111 (2006).

\bibitem{swendsen:2002} J.-S. Wang, R. H. Swendsen,
Journal of Statistical Physics, {\textbf 106}, pp 245-285 (2002).


\end{thebibliography}
\end{document}